\documentclass[longbibliography, aps, prl, amsmath, amssymb, amsfonts, twocolumn, superscriptaddress, footinbib, floatfix, 10pt]{revtex4-2}

\usepackage[german,american]{babel}
\usepackage{graphicx} 
\usepackage{dcolumn} 
\usepackage{bm} 
\usepackage{epsfig}
\usepackage{latexsym}

\usepackage[ruled, vlined, linesnumbered]{algorithm2e}
\usepackage{hyperref}
\usepackage[normalem]{ulem}
\usepackage{xcolor}
\usepackage{tikz}
\usetikzlibrary{arrows.meta,calc,positioning}
\usepackage{epstopdf}
\usepackage{mathtools} 
\usepackage{physics}
\usepackage{lipsum}
\usepackage{microtype}

\usepackage{enumitem}

\usepackage{amsthm}

\hypersetup{
	colorlinks,
	citecolor=blue,
	filecolor=blue,
	linkcolor=blue,
	urlcolor=blue
}

\usepackage[utf8]{inputenc}

\UseRawInputEncoding

\definecolor{darkblue}{rgb}{0,0,0.55}

\makeatletter
\let\frontmatter@footnote@produce\frontmatter@footnote@produce@footnote
\makeatother

\begin{document}

\title{Learning to Decode Concatenated Quantum Codes with Hierarchical Message Passing}

\author{Jiahui Wu}
\affiliation{Department of Physics, The Hong Kong University of Science and Technology, Clear Water Bay, Kowloon, Hong Kong SAR, China}

\author{Chao Zhang}
\affiliation{Department of Physics, The Hong Kong University of Science and Technology, Clear Water Bay, Kowloon, Hong Kong SAR, China}

\author{Zipeng Wu}
\affiliation{SpinQ Technology (Hong Kong) Co. Ltd., Cyberport, Pok Fu Lam, Hong Kong SAR, China}

\author{Shilin Huang}
\email{huangsl@ust.hk}
\affiliation{Department of Physics, The Hong Kong University of Science and Technology, Clear Water Bay, Kowloon, Hong Kong SAR, China}

\date{\today}

\begin{abstract}

	We introduce a neural message-passing framework for decoding general concatenated stabilizer codes. Soft beliefs propagate bidirectionally across concatenation levels, and lightweight neural networks learn only to aggregate incoming messages. For the concatenated $[[15,7,3]]$ quantum Hamming code, the resulting decoder achieves substantially higher thresholds than the state-of-the-art bidirectional hard-decision decoder under both bit-flip and depolarizing noise. In particular, the depolarizing pseudo-threshold nearly doubles, from $6.5\%$ to $12.3\%$. For many-hypercube codes, a decoder fine-tuned on circuit-level errors in Knill's teleportation-based error correction can achieve lower logical-CNOT failure rates than their dedicated decoder, using a fixed number of message-passing iterations instead of extensive combinatorial search. Our framework provides a generic decoding tool for exploring the design space of concatenated codes, including non-CSS constructions, toward low-overhead fault tolerance.

\end{abstract}

\maketitle

\textit{Introduction.}---Concatenated quantum codes recursively encode logical qubits into small- or medium-sized component codes~\cite{knill1996concatenated,nielsen2010quantum,rahn2002exact}. Their code distance grows exponentially with the number of concatenation levels. This recursive structure also provides a natural framework for logical gates and fault-tolerant gadgets. Historically, concatenated codes established the threshold theorem for fault-tolerant quantum computation~\cite{aharonov1997fault,knill1998resilient,aliferis2005quantum}. Knill subsequently showed that concatenating small error-detecting codes with error-correcting teleportation and post-selected ancilla preparation achieves fault-tolerance thresholds at the percent level~\cite{knill2005quantum}. 

Recently, high-rate concatenated codes have emerged as competitive candidates for low-overhead fault tolerance. Concatenating quantum Hamming codes with an increasing rate at each level achieves asymptotically efficient fault tolerance with constant space overhead~\cite{yamasaki2024time,yoshida2025concatenate}. Practically, recursive constructions based on small error-detecting codes attain high encoding rates and competitive thresholds under circuit-level noise~\cite{goto2024high,goto2025optimized}. Surface codes can also act as inner codes, with a high-rate outer code, so that local connectivity is achieved without sacrificing encoding rate~\cite{gidney2025yoked}.

Decoding high-rate concatenated codes efficiently and accurately remains difficult. Local hard-decision decoding, the simplest baseline, walks up the concatenation levels and settles on one error per block as it goes. A mistake at an early level is then permanent, so the decoder miscorrects low-weight errors and falls short of the code distance. Message-passing decoders attempt to retain this lost information~\cite{fern2008correctable,poulin2006optimal,goto2013soft,goto2024high}. However, passing the full error distribution up the hierarchy~\cite{poulin2006optimal} incurs a cost that grows exponentially with the number of logical qubits, while passing only single-qubit marginals~\cite{goto2024high} neglects critical intra-block correlations. Bidirectional hard-decision decoders pass higher-level constraints back down the hierarchy and revise lower-level commitments by combinatorial search, empirically reaching high thresholds while preserving the code distance~\cite{goto2024high,zhang2026bidirectional}. Yet these decoders must be custom-designed for each code family, and on difficult syndromes their combinatorial search can fail, or stall with unbounded worst-case latency.

In this work, we introduce a neural-augmented bidirectional message-passing decoder for concatenated quantum codes. To effectively exploit the hierarchical code structure, soft messages propagate both upward and downward across the concatenation levels. Unlike standard belief propagation~\cite{pearl1982reverend,mackay2004sparse}, which passes only scalar marginals, our decoder employs vector-valued beliefs to retain essential intra-block correlations. These messages are then combined using fixed arithmetic rules that are computationally simpler than standard BP updates. The neural augmentation is deliberately lightweight: small feedforward networks, shared across nodes at each level, are trained solely to aggregate messages arriving from different constraints. By keeping the core message-passing rules fixed and capping the maximum number of iterations, this architecture sharply departs from fully end-to-end neural decoders~\cite{torlai2017neural,krastanov2017deep,varsamopoulos2018decoding,ni2020neural,gicev2023scalable,sweke2021reinforcement,fitzek2020deep,bausch2024learning} and neural belief propagation schemes~\cite{liu2019neural,ninkovic2024decoding,maan2025machine} that learn the update rules themselves. Instead of predicting a recovery for each physical qubit, the decoder directly outputs the logical error class, preventing corrections that are inconsistent with the observed syndrome~\cite{chamberland2018deep,baireuther2018machine,bausch2024learning,gu2026cascade,zhang2026learning}. It executes a fixed number of message-passing steps per syndrome, inherently guaranteeing bounded worst-case latency.

We evaluate this architecture across three concatenated code families, training a separate decoder for each. On the $[[15,7,3]]$ quantum Hamming code, our approach outperforms the bidirectional hard-decision decoder~\cite{zhang2026bidirectional}, nearly doubling the depolarizing (pseudo-)threshold. Under circuit-level noise, our decoder also achieves lower failure rates on the $[[6,4,2]]$ many-hypercube code~\cite{goto2024high} than bidirectional hard-decision baselines relying on extensive combinatorial search. We further demonstrate its applicability to non-CSS structures on the concatenated $[[8,3,3]]$ code~\cite{PhysRevA.54.1862}. Together, these results demonstrate a compelling combination of high accuracy, broad generality, and the strictly bounded worst-case latency required for practical fault tolerance.
\label{sec:graph}

\textit{Hierarchical decoding graph.}---We consider concatenated codes built
from the same component code $[[n,k,d]]$ at every level, for convenience.
The construction extends naturally to level-dependent component codes. A
\emph{component block} is one copy of this code, and an \emph{$\ell$-level
code} is the full concatenated code with $\ell$ levels. Level $0$ is the
physical qubits. An $\ell$-level code has $k^\ell$ logical qubits. To build an
$(\ell{+}1)$-level code, take $n$ copies of an $\ell$-level code. For each of
the $k^\ell$ logical-qubit indices, collect the corresponding logical qubit
from every copy. Each resulting set of $n$ qubits is grouped by one
component block into $k$ level-$(\ell{+}1)$ qubits. The $(\ell{+}1)$-level
code therefore contains $k^\ell$ component blocks, all instances of the same
$[[n,k,d]]$, with isomorphic stabilizer groups.

A message-passing decoder requires a
sparse decoding graph, but flattening a concatenated code to the physical level
turns every high-level stabilizer into a high-weight check and creates dense
short cycles, which are known obstacles for iterative quantum decoding~\cite{mackay2004sparse,poulin2008iterative}. We therefore keep the code's hierarchy as the graph. Each
level-$\ell$ qubit $\lambda$ carries three variable nodes $X_\lambda^{(\ell)}$,
$Y_\lambda^{(\ell)}$, $Z_\lambda^{(\ell)}$, one per Pauli type, and we use the
same symbol for a node and the operator it represents. This three-node unit is
a \emph{triple}. For each level-$\ell$ Pauli node $v$ above level $0$, we fix
a low-weight implementation as a product over its support $\partial v$ of
level-$(\ell-1)$ Pauli nodes. For each triple, the three implementations are
chosen to be compatible: they multiply to the identity up to phase, so the $Y$
implementation is the product of the $X$ and $Z$ implementations rather than
differing from it by a stabilizer. Each level-$\ell$ stabilizer $s$ is
likewise a product over its support $\partial s$ of level-$(\ell-1)$ Pauli nodes.
The $k$ triples of a component block, one per logical qubit, are its
\emph{canonical} triples.

Each variable node $q$ carries an \emph{error bit}
$b_q\in\{0,1\}$, equal to $1$ if and only if $q$ anticommutes with the physical
error. Three families of parity constraints relate these bits. First, each
stabilizer $s$ with syndrome $\sigma_s$, measured or derived as the mod-$2$ sum
of the syndromes of its factors, imposes the stabilizer constraint
\[
c_s:\quad \sum_{u\in\partial s} b_u=\sigma_s \pmod 2.
\]
Rather than restricting to the $m$ independent generators, we include every nonzero element of the stabilizer group as a constraint, for $\tilde{m}$ stabilizer constraints per block. Second, every level-$\ell$ variable $v$ above level $0$ is the product of the
level-$(\ell-1)$ variables in its support $\partial v$, giving the hierarchical
constraint
\[
c_v:\quad b_v=\sum_{u\in\partial v} b_u \pmod 2.
\]
Third, since $Y_\lambda^{(\ell)}=X_\lambda^{(\ell)}Z_\lambda^{(\ell)}$ up to
phase, the product $X_\lambda^{(\ell)}Y_\lambda^{(\ell)}Z_\lambda^{(\ell)}$ is
proportional to the identity, so any Pauli error anticommutes with zero or two
of the three operators. This gives the Pauli compatibility constraint
\[
c_\lambda:\quad b_{X_\lambda^{(\ell)}}+b_{Y_\lambda^{(\ell)}}+b_{Z_\lambda^{(\ell)}}=0 \pmod 2.
\]

To improve decoding, we augment each block with auxiliary logical-observable
triples. Within a
component block, choose a subset $T$ of at least two of its $k$ canonical
logical qubits and form the products $\prod_{i\in T}X_i$ and
$\prod_{i\in T}Z_i$. Each product is a logical observable. Multiplying it by a
stabilizer of the block leaves its encoded action unchanged, so we dress it to
a minimum-weight representative. The two dressed operators, with their product
as the $Y$ operator up to phase, form an \emph{auxiliary logical-observable
triple}. These auxiliary logical-observable triples add low-weight parity
constraints and short message-passing paths without adding logical qubits.
Each component block therefore has $k'$ logical-observable triples: its $k$
canonical logical triples plus $k'-k$ auxiliary logical-observable triples.

To build the next level, fix an index and collect the corresponding $n$
level-$\ell$ logical-observable triples, one from each of the $n$ component
blocks. These $n$ triples inherit the $\tilde{m}$ stabilizer constraints of the
component code. Only the first $k''$ indices, with $k''\leq k'$, are used to construct the
next level. For each selected index, the $n$ triples are combined according to
the $k'$ chosen logical observables, giving $k'$ level-$(\ell+1)$
logical-observable triples. The remaining $k'-k''$
triples still participate in the stabilizer constraints but do not generate
higher-level triples.

\label{sec:neural}

\textit{Neural message passing.}---Unlike standard belief propagation which passes probabilistic messages, our decoder directly evaluates smooth versions of the three parity constraints on continuous, high-dimensional soft beliefs. For each variable node $q$, the decoder maintains a state vector $\mathbf h_q^{(t)}\in[0,1]^r$ as a soft stand-in for the error bit $b_q$, initialized near $\mathbf 0$. During each iteration $t$, every constraint $c$ sends a message $\mathbf{\Delta}_{c\to q}^{(t)}$ to each variable $q$ that appears in it. Computed synchronously from the states of the other variables at iteration $t-1$, $\mathbf{\Delta}_{c\to q}^{(t)}$ indicates how $q$ should change to satisfy the constraint $c$. To capture the modulo-$2$ nature of the parity checks while ensuring the operations remain fully differentiable for training, these messages are evaluated using the periodic function $\phi(x) = \sin^2(\pi x/2) = (1-\cos(\pi x))/2$, which is applied element-wise to vectors as $\phi(\mathbf{x}) \coloneqq (\phi(x_1), \dots, \phi(x_r))$. For each variable, incoming messages of the same type are averaged into a single accumulated channel, which is then used to infer the variable's updated state $\mathbf h_q^{(t)}$.
	
We group these messages into four channels: one from the Pauli compatibility constraint, one from the stabilizer constraint, and two from the hierarchical constraints, one upward and one downward.

The \emph{cross-Pauli channel} is obtained from the Pauli compatibility constraints $c_\lambda\colon b_{X_\lambda^{(\ell)}} + b_{Y_\lambda^{(\ell)}} + b_{Z_\lambda^{(\ell)}} = 0$ defined for each triple $\lambda$. Because there is only one such constraint per variable, for a target $q=P_\lambda^{(\ell)}$ ($P\in\{X,Y,Z\}$) this gives directly:
\[
	\mathbf m_{\mathrm{cross},q}^{(t)}
	\coloneqq
	\mathbf{\Delta}_{c_\lambda\to q}^{(t)}
	\coloneqq
	\phi\!\left(\sum_{R\in\{X,Y,Z\}\setminus\{P\}}
		\mathbf h_{R_\lambda^{(\ell)}}^{(t-1)}\right).
\]

The \emph{upward} and \emph{downward channels} are obtained from the hierarchical constraints $c_v\colon b_v = \sum_{u\in\partial v} b_u$, which are defined for each variable $v$ above level $0$ and read in two complementary directions. Read upward, $c_v$ infers the parent $v$ from its children $\partial v$, yielding the upward channel directly:
\[
	\mathbf m_{\uparrow,v}^{(t)}
	\coloneqq
	\mathbf{\Delta}_{c_v\to v}^{(t)}
	\coloneqq
	\phi\!\left(\sum_{u\in\partial v}
		\mathbf h_u^{(t-1)}\right).
\]
Read downward, $c_v$ infers a target child $q\in\partial v$ from the parent and the remaining children:
\[
	\mathbf{\Delta}_{c_v\to q}^{(t)}
	\coloneqq
	\phi\!\left(\mathbf h_{v}^{(t-1)}
		-\sum_{u\in\partial v\setminus\{q\}}
		\mathbf h_u^{(t-1)}\right).
\]
Averaging over all parent variables $v$ containing $q$ yields the downward channel $\mathbf m_{\downarrow,q}^{(t)} \coloneqq \frac{1}{|\{v: q \in \partial v\}|} \sum_{v \ni q} \mathbf{\Delta}_{c_v\to q}^{(t)}$.

The \emph{stabilizer channel} is obtained from the stabilizer constraints $c_s\colon \sum_{u\in\partial s} b_u = \sigma_s$ defined for each stabilizer $s$. The constraint infers a target variable $q\in\partial s$ from the syndrome and the remaining variables:
\[
	\mathbf{\Delta}_{c_s\to q}^{(t)}
	\coloneqq
	\phi\!\left(\sigma_s\mathbf 1_r
		-\sum_{u\in\partial s\setminus\{q\}}\mathbf h_u^{(t-1)}\right).
\]
Averaging over all stabilizers $s$ acting on $q$ yields the stabilizer channel $\mathbf m_{\mathrm{stab},q}^{(t)} \coloneqq \frac{1}{|\{s: q \in \partial s\}|} \sum_{s \ni q} \mathbf{\Delta}_{c_s\to q}^{(t)}$.

These four accumulated channels drive the belief update. For a variable $q$ at level $\ell$, concatenate the active channels into a single vector $\mathbf m_q^{(t)}\in[0,1]^{4r}$. Two level-specific networks act on it: the update network $\mathrm{FFN}_{\mathrm{update}}^{(\ell,t)}$ maps $\mathbf m_q^{(t)}$ to a candidate feature $\mathbf c_q^{(t)}\in[0,1]^{r}$, while the gate network $\mathrm{FFN}_{\mathrm{gate}}^{(\ell,t)}$ reads the per-channel squared differences $(\mathbf 1_4\otimes\mathbf h_q^{(t-1)}-\mathbf m_q^{(t)})^{\odot2}$ from the previous state $\mathbf h_q^{(t-1)}$ and returns a gate $\mathbf a_q^{(t)}\in[0,1]^{r}$. The new state is their gated combination,
\[
	\mathbf h_q^{(t)}
	\coloneqq
	\mathbf a_q^{(t)}\odot\mathbf c_q^{(t)}
	+
	(1-\mathbf a_q^{(t)})\odot\mathbf h_q^{(t-1)}.
\]
For variables at the boundaries of the hierarchy, certain channels naturally do not exist---for example, bottom-level variables ($\ell=0$) have no children to provide upward channels, and top-level variables ($\ell=L$) have no parents to provide downward channels. The neural networks simply omit these missing channels from their inputs.

Reading the squared discrepancy (denoted by elementwise squaring $^{\odot2}$) lets the gate modulate the update by how consistent the incoming messages are with the current state. Both $\mathrm{FFN}$s are standard multilayer perceptrons with a single ReLU hidden layer. In our experiments the internal variable state dimension is $r=8$; the update network $\mathrm{FFN}_{\mathrm{update}}^{(\ell,t)}$ has a hidden dimension of $80$ and outputs through $\phi$ to ensure valid predictions, while the gate network $\mathrm{FFN}_{\mathrm{gate}}^{(\ell,t)}$ has a hidden dimension of $8$ and outputs through a sigmoid to constrain the weights to $[0,1]$.

The ultimate task of the decoder is to infer $b_v$ for the top-level logical variables, which dictates whether the recovery operation should anticommute with the corresponding logical operators. The message passing runs $T_0=2$ warm-up blocks followed by a stack of $T_1=4$ recurrent blocks repeated $R$ times, for $T = T_0 + R T_1$ iterations in total; at inference $R=22$, giving $T \le 90$. The logical bits are then read out directly, with no additional learned readout network. To keep the model size fixed as the number of iterations grows, the update and gate networks are shared across recurrent periods, $\mathrm{FFN}^{(\ell,t)}\equiv\mathrm{FFN}^{(\ell,t+T_1)}$ for $t>T_0$. For each top-level variable $v$, the decoder projects the final beliefs over its support $\partial v$, applies the smooth parity map, and thresholds the mean of the resulting $r$ components,
\[
	\hat b_v
	= \frac{1}{r}\sum_{j=1}^{r}\phi\!\left(\sum_{u\in\partial v}\mathbf h_{u,j}^{(T)}\right),
\]
predicting $b_v=1$ when $\hat b_v>\tfrac12$. After each stack of $T_1$ recurrent blocks, the decoder likewise infers the physical error bits from the level-$0$ beliefs and, if these inferred physical errors reproduce the measured syndrome and agree with the inferred logical errors, terminates early before the full $T$ iterations are exhausted. The model is trained by minimizing the binary cross-entropy between the logical predictions and the true bits.

\label{sec:numerics}

\begin{figure*}[t]
	\centering
	\includegraphics[width=1.0\linewidth]{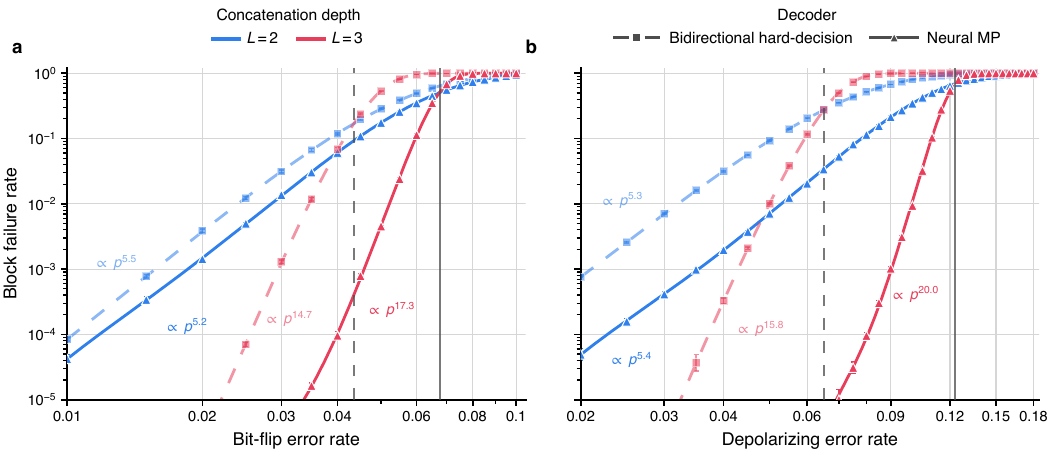}
	\caption{Block failure rate of the concatenated $[[15,7,3]]$ Hamming code at concatenation depths $L=2$ and $L=3$, comparing the neural message-passing (MP) decoder with the bidirectional hard-decision decoder. (a) Under bit-flip noise, the neural MP decoder raises the pseudo-threshold from $p^\ast\approx4.4\%$ to $6.8\%$. (b) Under depolarizing noise, it nearly doubles the pseudo-threshold, from $p^\ast\approx6.5\%$ to $12.3\%$.}
	\label{fig:code1573}
\end{figure*}

\textit{Numerical results.}---We test whether a single decoder architecture,
trained separately on each code, can outperform decoders designed for specific
code families. Specialized decoders exist for the concatenated $[[15,7,3]]$
quantum Hamming code~\cite{zhang2026bidirectional} and for
many-hypercube codes built from the $[[6,4,2]]$ code~\cite{goto2024high},
the latter of which additionally admits a circuit-level noise benchmark. For
other concatenated codes encoding multiple logical qubits, constructing a
hand-designed decoder beyond local hard-decision decoding is cumbersome,
especially for non-CSS codes. We therefore evaluate the neural message-passing (MP) decoder
on the concatenation of the non-CSS $[[8,3,3]]$ code~\cite{PhysRevA.54.1862}.
Across experiments the hierarchical message-passing structure
is shared; only the component code and the choice of logical-observable triples change,
and full details are given in the Supplemental
Material~\cite{supmat}. We quantify decoding quality by the \textit{block failure rate}, the
probability that decoding leaves a logical error on any logical qubit, and by
the \textit{pseudo-threshold} $p^\ast$, the physical error rate at which the
failure-rate curves of successive concatenation levels cross, so that below
$p^\ast$ an added level suppresses errors further.\footnote{We refer to this as a pseudo-threshold because under near-optimal decoding, logical error rates can initially rise before falling~\cite{poulin2006optimal}. Crossings at low concatenation levels thus represent a finite-size effect and may not coincide with the true asymptotic threshold.} 

Fig.~\ref{fig:code1573} compares the neural MP decoder with the bidirectional hard-decision decoder~\cite{zhang2026bidirectional} on the concatenated $[[15,7,3]]$ quantum Hamming code~\cite{PhysRevA.54.1098,PhysRevA.54.4741}, which at level $L$ encodes $k=7^L$ logical qubits into $n=15^L$ physical qubits. Each block carries $7$ canonical logical triples plus $35$ auxiliary weight-$3$ logical-observable triples, giving $k'=42$; $k''=14$ of these are used to generate logical-observable triples at the next level. As shown in Fig.~\ref{fig:code1573}(a), under bit-flip noise, the neural MP decoder raises the pseudo-threshold from $p^\ast\approx4.4\%$ to $p^\ast\approx6.8\%$, and at $L=3$ its block failure rate is two to three orders of magnitude lower than the bidirectional decoder's, reaching $1.6\times10^{-5}$ versus $1.2\times10^{-2}$ at $p=0.035$. The bidirectional decoder empirically preserves the code distance but bounds its search space to keep the worst-case runtime manageable, and so fails on complex syndromes. The neural MP decoder instead learns to decode these syndromes directly. As shown in Fig.~\ref{fig:code1573}(b), under depolarizing noise, the improvement is larger: the neural MP decoder nearly doubles the pseudo-threshold, from $p^\ast\approx6.5\%$ to $p^\ast\approx12.3\%$, and at $p=0.07$ its failure rate is at least four orders of magnitude lower. The bidirectional decoder treats the $X$ and $Z$ syndromes separately and cannot exploit the $X$--$Z$ correlations of depolarizing noise, whereas the neural MP decoder couples them through the cross-Pauli channel.

\begin{figure}[t]
	\centering
	\includegraphics[width=1.0\linewidth]{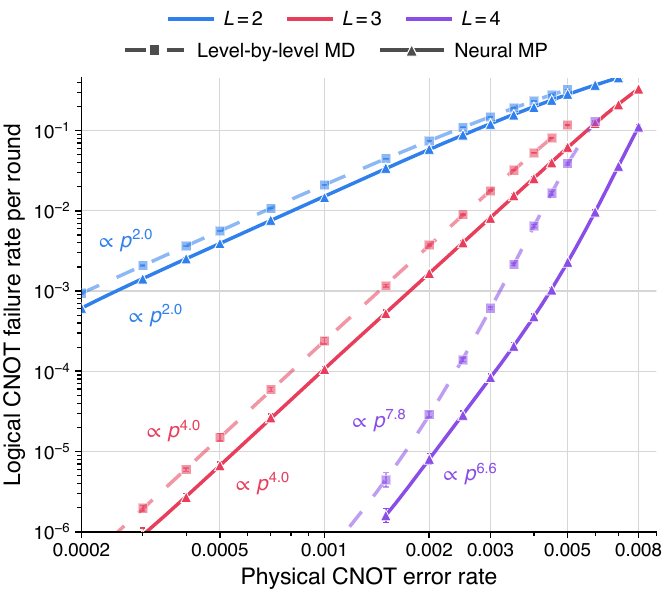}
	\caption{Evaluation of the neural message-passing (MP) decoder and the level-by-level minimum-distance (MD) decoder at concatenation depths $L=2$,$3$ and $4$ for the many-hypercube code family in circuit-level simulations under depolarizing noise.}
	\label{fig:codeMHC6}
\end{figure}

We also show that the decoder can be fine-tuned to circuit-level noise, using the many-hypercube code built from the $[[6,4,2]]$ code. At level $L$ it encodes $k=4^L$ logical qubits into $n=6^L$ physical qubits; each block carries $k'=15$ logical-observable triples, the four canonical weight-$2$ pairs plus their stabilizer-dressed products. Of these, $k''=9$ are used to generate logical-observable triples at the next level. Decoding is performed under Knill error-correction teleportation with postselected logical-zero preparation. We consider a circuit-level depolarizing noise model where preparations, measurements, and two-qubit gates fail with probability $p$, while single-qubit gates and idle qubits are assumed noiseless. Consequently, the residual errors are location-dependent and correlated rather than i.i.d. The full circuit construction, noise model, and fine-tuning procedure are detailed in the Supplemental Material~\cite{supmat}. We report the per-round logical-CNOT block failure rate $P_1=1-(1-P_{10})^{1/10}$, where $P_{10}$ is the failure probability over ten protected rounds. By learning the correlated, position-dependent error biases induced by the preparation circuits, the fine-tuned neural MP decoder consistently outperforms the level-by-level minimum-distance decoder of Ref.~\cite{goto2024high}. For example, at $L=4$ and $p=0.005$, the neural MP decoder achieves a block failure rate 17 times lower than that of Ref.~\cite{goto2024high}. As $p$ decreases, this relative improvement narrows because both decoders begin to hit an underlying error floor; at $p=0.002$, the block failure rate is 3.6 times lower. This residual floor arises because the level-$4$ state-preparation circuit is not strictly fault tolerant, meaning a weight-$6$ preparation error can already fail a block regardless of the decoder used.

\begin{figure}[t]
	\centering
	\includegraphics[width=1.0\linewidth]{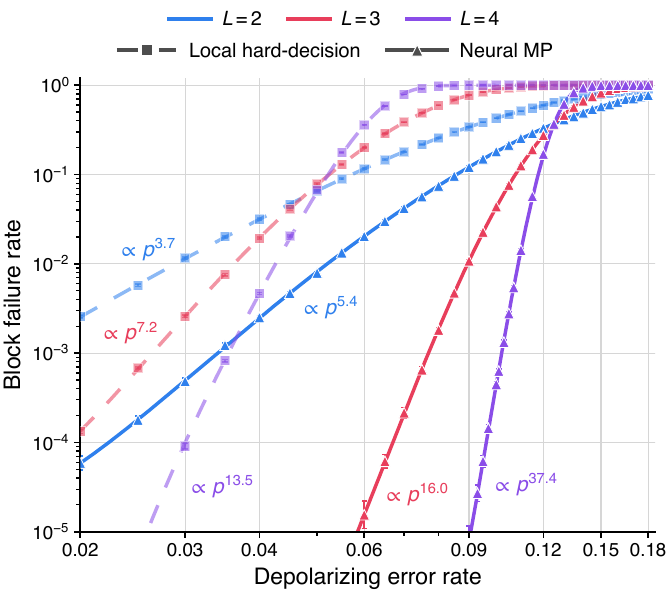}
	\caption{Evaluation of the neural message-passing (MP) decoder and the local hard-decision decoder at concatenation depths $L=2,3$ and $4$ for the non-CSS concatenated $[[8,3,3]]$ code family under depolarizing noise.}
	\label{fig:code833}
\end{figure}

Our decoder also achieves much lower failure rates on non-CSS concatenated codes under depolarizing noise, where local hard-decision decoding is the best available hard decoder but cannot preserve the code distance. The $[[8,3,3]]$ code encodes $k=3^L$ logical qubits into $n=8^L$ physical qubits at level $L$; its stabilizers are not purely $X$- or $Z$-type, and all $k'=k''=7$ weight-$4$ logical-observable triples are used to generate logical-observable triples at the next level. As shown in Fig.~\ref{fig:code833}, the neural MP decoder raises the pseudo-threshold from $p^\ast\approx5\%$ to $p^\ast\approx12.5\%$, and at every level its sub-threshold failure rate falls markedly more steeply with decreasing $p$, reflecting the effective distance lost by local hard-decision decoding.

\label{sec:conclusion}

\textit{Conclusion.}---We have introduced a hierarchical neural message-passing framework for decoding concatenated quantum codes. The message-passing rules are code-independent, and only the message combination and belief update are learned. The framework is generic, requiring per code only the choice of logical-observable triples of the component code that generate the hierarchical decoding graph. The trained decoders outperform the specialized hard-decision decoders built for those code families. On CSS codes the framework couples the $X$ and $Z$ syndromes, exploiting the correlations of depolarizing noise. The framework extends to non-CSS codes. Fine-tuning adapts a pretrained decoder to circuit-level noise. The decoder runs at most a fixed number of message-passing steps, so the worst-case decoding latency is bounded.

Our results establish a clear pathway toward realizing the practical potential of concatenated codes. The substantial improvements we demonstrate in decoding performance provide a strong mandate for optimizing fault-tolerant circuits and co-designing them with the decoder. From a theoretical perspective, while the pseudo-thresholds we observe still fall short of the zero-rate hashing bounds of $p \approx 18.9\%$ for depolarizing noise~\cite{bennett1996mixed} and $p \approx 11.0\%$ for bit-flip noise~\cite{PhysRevA.54.1098}, this gap is not necessarily fundamental. Because logical error rates can initially rise before falling under optimal decoding~\cite{poulin2006optimal}, crossings observed at currently accessible concatenation levels may not represent the true asymptotic threshold. Our present data cannot definitively disentangle such finite-size effects from inherent decoder limitations. Developing simpler, highly scalable decoding architectures will allow us to track these crossings at deeper levels and ultimately settle this open question.

This work is supported under National Key Research and Development Project of China (Grant No. 2025YFE0200900).

\bibliography{ref}

\clearpage

\onecolumngrid

\begin{center}
{\large\bfseries Supplemental Material: Learning to Decode Concatenated Quantum Codes with Hierarchical Message Passing}
\end{center}

\setcounter{section}{0}
\setcounter{equation}{0}
\setcounter{figure}{0}
\setcounter{table}{0}
\setcounter{secnumdepth}{2}
\renewcommand{\thesection}{S\arabic{section}}
\renewcommand{\thesubsection}{\Alph{subsection}}
\renewcommand{\theequation}{S\arabic{equation}}
\renewcommand{\thefigure}{S\arabic{figure}}
\renewcommand{\thetable}{S\arabic{table}}

\section{Component-Code Supports and Logical-Observable Triples}
\label{app:redundant-nodes}

We adopt the notation of the hierarchical decoding graph described above. Each
component block of $[[n,k,d]]$ carries $k'$ logical-observable triples: its $k$
canonical logical triples plus $k'-k$ auxiliary logical-observable triples, the
latter stabilizer-dressed to minimum weight. Of these, $k''$ triples form node
triples at the next level;
the remaining $k'-k''$ are local, generating stabilizer constraints but no node triples. Each block
carries its $m$ independent stabilizer generators together with $\tilde{m}-m$
redundant stabilizer constraints (products of the independent generators), so that
$\tilde{m}$ stabilizer constraint nodes per block enter the graph.

\begin{table}[htbp]
\centering
\caption{Component-code parameters: $k'$ logical-observable triples per block,
$k''$ node triples, $m$ independent stabilizer generators, and $\tilde{m}$
stabilizer constraint nodes per block.}
\label{tab:component-code-parameters}
\begin{tabular}{lccccccc}
\hline
code & $n$ & $k$ & $d$ & $k'$ & $k''$ & $m$ & $\tilde{m}$ \\
\hline
$[[15,7,3]]$ & 15 & 7 & 3 & 42 & 14 & 8 & 45 \\
$[[8,3,3]]$ & 8 & 3 & 3 & 7 & 7 & 5 & 31 \\
$[[6,4,2]]$ & 6 & 4 & 2 & 15 & 9 & 2 & 3 \\
\hline
\end{tabular}
\end{table}

\subsection{Concatenated \texorpdfstring{$[[15,7,3]]$}{[[15,7,3]]} quantum Hamming codes}

The $[[15,7,3]]$ quantum Hamming code \cite{PhysRevA.54.1098, PhysRevA.54.4741}
is the self-dual CSS code built from the classical $[15,11,3]$ Hamming code,
with $m=8$ independent generators read from the $4\times15$ parity-check matrix.
Logical-$Z$ and logical-$Y$ use the same physical support as logical-$X$ by
self-duality. For every $3$-element subset $T\subseteq\{1,\ldots,7\}$, the
product $\prod_{i\in T}\overline P_i$ has a weight-$3$ representative after
stabilizer dressing; the $\binom{7}{3}=35$ such subsets form the Steiner triple
system $STS(15)$. We form node triples at the next level from $k''=14$ triples: the $7$ canonical triples plus
the $7$ weight-$3$ auxiliary logical-observable triples below, whose $7\times7$ incidence matrix
has full rank over $\mathbb F_2$. The remaining $28$ auxiliary logical-observable triples are local.

In addition to the $m=8$ independent generators, each block carries redundant
stabilizer constraints obtained as binary products of these generators. Since the
$[[15,7,3]]$ code is self-dual, every nontrivial product again has weight $8$,
so we retain all $2^4-1=15$ of them---the $4$ generators, their $6$ pairwise
products, $4$ triple products, and the single four-fold product---in each of
the $X$, $Y$, and $Z$ sectors, giving $\tilde{m}=45$ stabilizer constraint nodes per block
($15$ per Pauli sector). Each redundant syndrome is the mod-$2$ sum of the
corresponding generator syndromes and requires no additional measurement.

\begin{table}[htbp]
\centering
\caption{Canonical logical-$X$ supports (left) and the seven selected
weight-$3$ auxiliary logical-$X$ representatives (right) of the
$[[15,7,3]]$ code. Auxiliary entries are written as
$\prod_{i\in T}\overline X_i\equiv R$ up to an $X$-type stabilizer.}
\label{tab:1573-logicals}
\begin{tabular}{@{}ll@{\qquad}ll@{}}
\hline
canonical & support & auxiliary & representative \\
\hline
$\overline X_1$ & $X_1X_2X_4X_8X_{15}$ & $\overline X_1\overline X_2\overline X_3$ & $X_4X_{10}X_{14}$ \\
$\overline X_2$ & $X_1X_2X_5X_{10}X_{12}$ & $\overline X_1\overline X_2\overline X_6$ & $X_3X_{12}X_{15}$ \\
$\overline X_3$ & $X_1X_2X_6X_{11}X_{14}$ & $\overline X_1\overline X_3\overline X_7$ & $X_3X_8X_{11}$ \\
$\overline X_4$ & $X_1X_2X_7X_9X_{13}$ & $\overline X_2\overline X_3\overline X_5$ & $X_3X_5X_6$ \\
$\overline X_5$ & $X_1X_4X_6X_9X_{10}$ & $\overline X_4\overline X_5\overline X_6$ & $X_5X_9X_{12}$ \\
$\overline X_6$ & $X_1X_4X_7X_{12}X_{14}$ & $\overline X_4\overline X_5\overline X_7$ & $X_6X_{11}X_{13}$ \\
$\overline X_7$ & $X_1X_8X_{10}X_{13}X_{14}$ & $\overline X_4\overline X_6\overline X_7$ & $X_7X_8X_{15}$ \\
\hline
\end{tabular}
\end{table}

\subsection{Concatenated non-CSS \texorpdfstring{$[[8,3,3]]$}{[[8,3,3]]} codes}

The non-CSS $[[8,3,3]]$ code \cite{PhysRevA.54.1862} has $m=5$ independent stabilizer generators,
\begin{align}
    S_1 &= X_1X_2X_3X_4X_5X_6X_7X_8, \nonumber\\
    S_2 &= Z_1Z_2Z_3Z_4Z_5Z_6Z_7Z_8, \nonumber\\
    S_3 &= Z_3Y_4X_5Z_6Y_7X_8, \nonumber\\
    S_4 &= Z_2X_3X_5Y_6Z_7Y_8, \nonumber\\
    S_5 &= X_2Z_4Z_5X_6Y_7Y_8.
    \label{eq:833-generators}
\end{align}
The canonical logical basis is \cite{chen2025fault}
\begin{align}
    \overline X_1 &= X_4X_5X_7X_8,
    & \overline Z_1 &= Z_2X_3Z_5X_8,
    & \overline Y_1 &= Z_2X_3X_4Y_5X_7, \nonumber\\
    \overline X_2 &= X_2X_3Y_7Y_8,
    & \overline Z_2 &= Z_2Z_3Z_4Z_8,
    & \overline Y_2 &= Y_2Y_3Z_4Y_7X_8, \nonumber\\
    \overline X_3 &= Z_1Z_2X_6X_7,
    & \overline Z_3 &= Z_1Z_2Z_4Z_7,
    & \overline Y_3 &= Z_4X_6Y_7.
    \label{eq:833-canonical-logicals}
\end{align}
In addition to the $m=5$ generators of Eq.~\eqref{eq:833-generators}, each
block carries redundant stabilizer constraints obtained as nonidentity products of
these generators. We retain every product whose physical weight is at most $8$,
namely all $2^5-1=31$ nonidentity stabilizers---the three weight-$8$ elements
and the $28$ weight-$6$ elements---so each block carries $\tilde{m}=31$ stabilizer
constraint nodes. Each redundant syndrome is the mod-$2$ sum of the corresponding
generator syndromes.
All $k'=7$ triples form node triples at the next level ($k''=k'$), so there are no local triples. The four auxiliary logical-observable triples are stabilizer-dressed to the weight-$4$ minimum of their
cosets:

\begin{table}[htbp]
\centering
\caption{Auxiliary logical-$X$, logical-$Z$, and logical-$Y$ representatives
of the $[[8,3,3]]$ code.}
\label{tab:833-auxiliary-logicals}
\begin{tabular}{lll}
\hline
$\overline X_4 = X_1X_2Z_4Z_5$ & $\overline Z_4 = X_1Z_4Z_6X_7$ & $\overline Y_4 = X_2Z_5Z_6X_7$ \\
$\overline X_5 = Y_1Z_4Y_5Z_6$ & $\overline Z_5 = Y_1X_4X_5Y_8$ & $\overline Y_5 = Y_4Z_5Z_6Y_8$ \\
$\overline X_6 = Y_3Z_6Y_7Z_8$ & $\overline Z_6 = Y_2Y_6Y_7Y_8$ & $\overline Y_6 = Y_2Y_3X_6X_8$ \\
$\overline X_7 = Y_1X_3X_5Y_6$ & $\overline Z_7 = Z_1Z_3Y_5Y_6$ & $\overline Y_7 = X_1Y_3Z_5$ \\
\hline
\end{tabular}
\end{table}

\subsection{Many-hypercube codes}

The $[[6,4,2]]$ block has $m=2$ independent stabilizer generators, $S_X=X_1\cdots X_6$ and $S_Z=Z_1\cdots Z_6$. In addition to these two generators, each block carries one redundant stabilizer constraint, the product $S_Y=S_XS_Z=Y_1\cdots Y_6$ (up to phase), whose syndrome is the mod-$2$ sum of the two generator syndromes; this gives $\tilde{m}=3$ stabilizer constraint nodes per block. Products of its four canonical logical pairs, stabilizer-dressed when needed, generate the $2^4-1=15$ nonidentity logical-observable triples in each Pauli sector. We form node triples at the next level from $k''=9$ triples: the four canonical triples, the product of all four, and the four pairwise products $\{1,2\}$, $\{3,4\}$, $\{1,3\}$, $\{2,4\}$; the remaining $6$ triples are local.

\begin{table}[htbp]
\centering
\caption{Canonical logical-$X$, logical-$Z$, and logical-$Y$ pairs of the
$[[6,4,2]]$ code.}
\label{tab:642-canonical-logicals}
\begin{tabular}{lll}
\hline
$\overline X_1 = X_1X_2$ & $\overline Z_1 = Z_2Z_6$ & $\overline Y_1 = X_1Y_2Z_6$ \\
$\overline X_2 = X_1X_3$ & $\overline Z_2 = Z_3Z_6$ & $\overline Y_2 = X_1Y_3Z_6$ \\
$\overline X_3 = X_1X_4$ & $\overline Z_3 = Z_4Z_6$ & $\overline Y_3 = X_1Y_4Z_6$ \\
$\overline X_4 = X_1X_5$ & $\overline Z_4 = Z_5Z_6$ & $\overline Y_4 = X_1Y_5Z_6$ \\
\hline
\end{tabular}
\end{table}

\begin{table}[htbp]
\centering
\caption{The five selected auxiliary logical-$X$, logical-$Z$, and logical-$Y$ representatives of the $[[6,4,2]]$ code.}
\label{tab:642-auxiliary-logicals}
\begin{tabular}{lll}
\hline
$\overline X_5 = X_1X_6$ & $\overline Z_5 = Z_1Z_6$ & $\overline Y_5 = Y_1Y_6$ \\
$\overline X_6 = X_2X_3$ & $\overline Z_6 = Z_2Z_3$ & $\overline Y_6 = Y_2Y_3$ \\
$\overline X_7 = X_4X_5$ & $\overline Z_7 = Z_4Z_5$ & $\overline Y_7 = Y_4Y_5$ \\
$\overline X_8 = X_2X_4$ & $\overline Z_8 = Z_2Z_4$ & $\overline Y_8 = Y_2Y_4$ \\
$\overline X_9 = X_3X_5$ & $\overline Z_9 = Z_3Z_5$ & $\overline Y_9 = Y_3Y_5$ \\
\hline
\end{tabular}
\end{table}
\section{Neural Decoding with Hierarchical Message Passing}
\label{app:decoder}

The preceding section fixes the hierarchical graph. Here we describe how training samples and exact targets are generated on that graph, how boundary channels and the readout are implemented, and how the decoder is optimized and selected. The message definitions and node-update rules follow the notation of the main text.

With qubit addresses $\mathcal Q_\ell$ at each level $\ell$, the set of variable nodes is
\begin{equation}
    \mathcal V_\ell=
    \{P_\lambda^{(\ell)}:\lambda\in\mathcal Q_\ell,\,P\in\mathsf P = \{X,Y,Z\}\},
\end{equation}
so $|\mathcal V_\ell|=3|\mathcal Q_\ell|$. A top-level logical variable $v=P_\lambda^{(L)}\in\mathcal V_L$ is the Pauli operator $P$ acting on the logical qubit at address $\lambda\in\mathcal Q_L$. The canonical top-level logical variables are used for decoding-failure reporting; the auxiliary top-level logical-observable variables, when supervised, are deterministic parities of the canonical variables and add no encoded degrees of freedom. Local triples produce no node triples and are not expanded into top-level prediction targets.

\subsection{Training samples and exact targets}

Training data are generated online under an independent depolarizing channel. For every sample, a physical error rate $p$ is first drawn uniformly from $[0,p_{\max}]$, after which the Pauli error $E=\bigotimes_{\mu=1}^{n}E_\mu$ is sampled according to
\begin{equation}
    \begin{aligned}
        \Pr(E_\mu=I)&=1-p,
        \\
        \Pr(E_\mu=X)&=\Pr(E_\mu=Y)=\Pr(E_\mu=Z)=\frac{p}{3}.
    \end{aligned}
    \label{eq:training-depolarizing-channel}
\end{equation}

The exact target of a variable node $q$ is its error bit $b_q\in\{0,1\}$ of the main text, namely $b_q=1$ exactly when the sampled error $E$ anticommutes with the Pauli operator represented by $q$. The exact bits of higher-level variables and the syndrome bits are obtained recursively from the hierarchical and stabilizer constraints of the main text. Auxiliary top-level logical-observable variables provide deterministic targets, not additional measurements or encoded degrees of freedom.

\subsection{State initialization and boundary channels}

Each variable node $q$ carries a state vector $\mathbf h_q^{(t)}\in[0,1]^r$ after message-passing iteration $t$, a learned soft feature associated with its error bit and not assumed to be a calibrated probability. In the code-capacity experiments the initial state is sampled independently as
\begin{equation}
    \mathbf h_q^{(0)}\sim \operatorname{Unif}([0,0.01]^r).
    \label{eq:initial-latent-state}
\end{equation}

The four accumulated channels of the main text are concatenated in the implementation order $(\uparrow,\downarrow,\mathrm{stab},\mathrm{cross})$ into a single $4r$-dimensional update input. At the two hierarchy boundaries one ordinary hierarchical channel is absent: a physical-level node $q=P_\lambda^{(0)}\in\mathcal V_0$ has no lower-level support and thus no upward channel, and a top internal node $q=P_\lambda^{(L-1)}\in\mathcal V_{L-1}$ has no parent and thus no downward channel. In both cases the absent channel is zero-padded so the update network always receives the full $4r$-dimensional input.

\subsection{Readout, loss, and failure reporting}

The logical readout is defined in the main text: after $T$ iterations, the final beliefs of each top-level logical variable $v\in\mathcal V_L$ are projected over its support $\partial v$, passed through the smooth parity map, and thresholded. During training an auxiliary physical readout is used, defined as the mean of the $r$ components of the final level-$0$ state,
\begin{equation}
    \hat b_q
    = \frac{1}{r}\sum_{j=1}^{r}\mathbf h_{q,j}^{(T)},
    \label{eq:physical-readout-probability}
\end{equation}
which lies in $[0,1]$. The decoder is trained end-to-end with binary cross-entropy at the physical qubits and the top-level logical qubits, evaluated directly on these scalars without logits or a sigmoid. For one sample the physical and logical losses are
\begin{equation}
    \begin{aligned}
    \mathcal L_{\mathrm{phy}}
    &=-\frac{1}{|\mathcal V_0|}
    \sum_{q\in\mathcal V_0}
    \left[b_q\ln \hat b_q
        +(1-b_q)\ln(1-\hat b_q)\right],
    \label{eq:physical-bce-loss}\\
    \mathcal L_{\mathrm{log}}
    &=-\frac{1}{|\mathcal V_L|}
    \sum_{v\in\mathcal V_L}
    \left[b_v\ln \hat b_v
        +(1-b_v)\ln(1-\hat b_v)\right],
    \end{aligned}
\end{equation}
with total loss $\mathcal{L}=\mathcal{L}_{\mathrm{phy}}+\mathcal{L}_{\mathrm{log}}$. Supervising the auxiliary logical-observable variables in $\mathcal V_L$ does not change the encoded dimension: logical failure reporting and validation use only the independent logical variables. Since the target bits of the Pauli-$Y$ variables are determined by the corresponding Pauli-$X$ and Pauli-$Z$ bits, failure reporting uses only the Pauli-$X$ and Pauli-$Z$ bits on the independent logical qubits. For a top-level logical variable $v\in\mathcal V_L$ the decoder predicts the logical bit $\hat l_v=\mathbf 1[\hat b_v>1/2]$, and a decoding failure is reported if $\hat l_v\neq b_v$ for any canonical top-level logical Pauli-$X$ or Pauli-$Z$ variable $v\in\mathcal V_L$.

\subsection{Training schedule and hyperparameters}

The recurrent stack of the main text (with $T_0=2$ warm-up blocks and $T_1=4$ recurrent blocks) is repeated $R=5$ times during training, giving $T=T_0+RT_1=22$ message-passing blocks. The last three recurrent outputs, after $t\in\{14,18,22\}$ blocks, are supervised:
\begin{equation}
    \begin{aligned}
        \mathcal L_{\mathrm{train}}
        &=\frac{1}{3} \left(\mathcal L^{(14)}+
            \mathcal L^{(18)}+
            \mathcal L^{(22)}\right),
        \\
        \mathcal L^{(t)}
        &=\mathcal L_{\mathrm{phy}}^{(t)}
        +\mathcal L_{\mathrm{log}}^{(t)}.
    \end{aligned}
    \label{eq:late-iteration-training-loss}
\end{equation}
Multiple stabilizer and downward messages arriving at a variable node are aggregated by average pooling, as in the main text, with a message dropout of $0.05$ applied to the stabilizer and higher-level logical messages for training stability and expressive generalization. Training runs for $500$ epochs with a fixed budget of $16000$ freshly sampled syndromes per epoch, split into batches according to Table~\ref{tab:neural-training-batch-sizes}. Two-dimensional matrix weights of the feedforward networks are optimized by Muon with learning rate $10^{-3}$; biases, embeddings, gates, and other parameters are optimized by AdamW with learning rate $10^{-4}$, one tenth of the Muon learning rate.

\begin{table}[htbp]
    \centering
    \caption{Code-family-specific training batch sizes at each concatenation level.}
    \label{tab:neural-training-batch-sizes}
    \begin{ruledtabular}
    \begin{tabular}{lccc}
        Component code & $[[15,7,3]]$ & $[[8,3,3]]$ &  $[[6,4,2]]$  \\
        \hline
        Batch size (Lv2) & $32$ & $32$ & $32$\\
        Batch size (Lv3) & $24$ & $32$ & $32$\\
        Batch size (Lv4) & n/a & $32$ & $32$\\
    \end{tabular}
    \end{ruledtabular}
\end{table}

The physical error rates used for training are sampled uniformly from $[0,p_{\max}]$, where $p_{\max}=0.14$ for concatenated $[[15,7,3]]$ and $[[8,3,3]]$ codes and $p_{\max}=0.09$ for many-hypercube codes. The validation physical error rates are fixed at $p=0.10$ for concatenated $[[15,7,3]]$ and $[[8,3,3]]$ codes and $p=0.07$ for many-hypercube codes. Validation is performed after every epoch using $R=7$ recurrent repetitions, corresponding to $T=T_0+RT_1=30$ message-passing blocks. The checkpoint with the lowest validation block failure rate is selected as the final model.
\section{Circuit-Level Simulation for Many-Hypercube Codes}
\label{app:circuit}

Code-capacity benchmarks sample an independent Pauli error directly on the physical data qubits and then decode with ideal syndrome information. Circuit-level benchmarks test a more structured noise source: errors are generated by faulty state preparations, gates, and measurements inside a concrete fault-tolerant logical operation, and they propagate through multi-qubit gates such as CNOTs. The resulting error on the data qubits is therefore correlated and position-dependent rather than independent, so the circuit-level benchmark is a more realistic test of a decoder for fault-tolerant quantum computing. We outline here the circuit-level simulation used for the many-hypercube codes, following Ref.~\cite{goto2024high}, whose error-correction mechanism is Knill's error correction by teleportation (ECT)~\cite{knill2005quantum,knill2005scalable}.

\subsection{Circuit-Level Error Model}
\label{sec:circ-noise}

State preparation, measurement, and each gate fail independently with the same depolarizing error probability $p$: a zero-state preparation introduces an $X$ error with probability $p$, a measurement outcome is flipped with probability $p$, and each CNOT is followed by one of the fifteen nonidentity two-qubit Pauli errors with probability $p$, so that each individual two-qubit error occurs with probability $p/15$. The single-qubit Hadamard gate is taken to be noiseless. Pauli errors propagate through the circuit: a Hadamard exchanges $X$ and $Z$, while a CNOT copies an $X$ error from its control to its target and a $Z$ error from its target to its control. A single faulty gate can therefore create a correlated multi-qubit error on the data qubits. The simulation tracks the full propagated error $E$ on the data qubits explicitly, so that the exact syndrome and logical error are available at every step.

\subsection{Error Correction by Teleportation}
\label{sec:circ-ect}

\begin{figure}[!htbp]
    \centering
    \includegraphics[width=\linewidth]{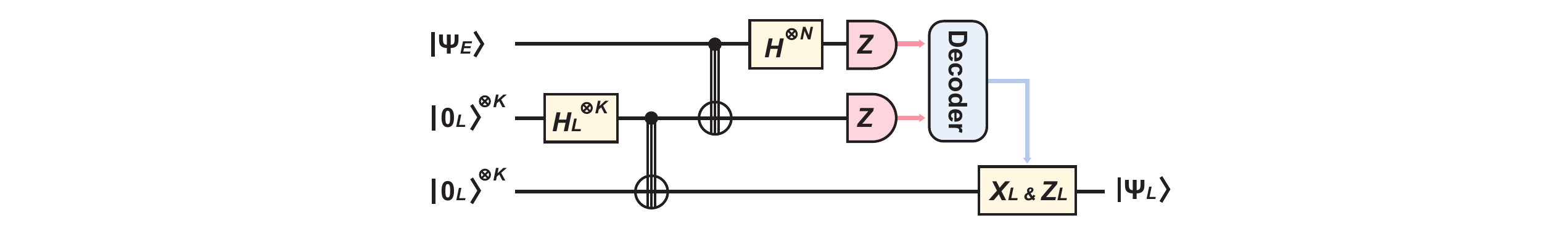}
    \caption{Knill error correction by teleportation for an encoded block containing $K$ logical qubits. The input block $|\Psi_E\rangle$ is coupled transversally to the first half of an encoded Bell pair. The input and first Bell-half blocks are measured in the $X$ and $Z$ bases, respectively; a decoder infers the logical Pauli byproduct, and the corresponding $X_L$ and $Z_L$ corrections are applied to the surviving Bell half to produce $|\Psi_L\rangle$.}
    \label{fig:knill-ect}
\end{figure}

For a Pauli error $E$ on the $n$ data qubits, let $b_q(E)$ denote the error bit defined in the main text, where $b_q(E)=1$ if and only if $E$ anticommutes with the Pauli operator represented by the variable node $q$. At the physical level, the $n$ bits $b_q(E)$ at the $X$-type nodes and the $n$ bits at the $Z$-type nodes form the binary symplectic representation of $E$---a $2n$-bit vector over $\mathbb F_2$ used to identify $E$. The syndrome $\sigma(E)$ collects the stabilizer syndrome bits $\sigma_s=\sum_{q\in\partial s}b_q(E)$ over all stabilizers $s$, while the logical error class $L(E)$ collects the error bits $b_v(E)$ for the canonical top-level logical variables $v\in\mathcal V_L$. These variables consist entirely of logical $X$- and $Z$-type variables; the logical $Y$-type bit at each address is simply the mod-$2$ sum of the two and carries no independent information. Consequently, $L(E)$ is exactly the logical Pauli coset of $E$. Both $\sigma$ and $L$ are linear maps over $\mathbb F_2$ that together specify the decoding task completely: the decoder receives the syndrome $\sigma(E)$ as input and must return the logical error class $L(E)$.

As shown in Fig.~\ref{fig:knill-ect}, Knill's ECT processes an input logical block by passing it through an encoded Bell pair (whose fault-tolerant preparation is described in Sec.~\ref{sec:circ-state-prep}). An encoded block carries $K$ logical qubits, and the encoded Bell pair over two such blocks is
\begin{equation}
    |\mathrm{Bell}_L\rangle
    =\left(\frac{|0_L\rangle|0_L\rangle+|1_L\rangle|1_L\rangle}{\sqrt{2}}\right)^{\otimes K},
\end{equation}
where $|0_L\rangle$ and $|1_L\rangle$ denote the two logical computational-basis states of a single logical qubit. A transversal logical CNOT couples the input block to the first half of the Bell pair. The input block is then measured destructively in the $X$ basis, while the first Bell half is measured in the $Z$ basis. Finally, the logical Pauli correction inferred from these two outcomes is applied as a byproduct correction to the surviving second Bell half---the teleported output block.

Suppose first that the ECT circuit itself is error-free. The $X$-basis measurement of the input block yields the $n$ bits of the $Z$ component of a Pauli operator, and the $Z$-basis measurement of the first Bell half yields the $n$ bits of its $X$ component. Together, these two outcome strings form the binary symplectic representation of a single $n$-qubit Pauli operator $M$, following the same format used for $E$. The ideal outcome $M$ is uniformly random with a vanishing syndrome ($\sigma(M)=0$). Its logical class $L(M)$ is exactly the byproduct correction to be applied to the teleported output block, requiring no further decoding. 

If the input block instead carries a Pauli error $E$, the recorded outcome becomes $M'=M\oplus E$. By linearity, $\sigma(M')=\sigma(M)\oplus\sigma(E)=\sigma(E)$, meaning the decoder is presented with the syndrome of $E$ alone. Whenever the decoder infers the logical class $L(E)$ correctly from this syndrome, the required correction is recovered as
\begin{equation}
    L(M)=L(M')\oplus L(E),
    \label{eq:circ-ect-two-operator}
\end{equation}
which is the sum of a byproduct component $L(M')$ read directly from the noisy outcomes and a decoded component $L(E)$. This two-component decomposition accurately reflects how the training data in Sec.~\ref{sec:circ-dataset} are generated. If the decoder mispredicts $L(E)$, the applied correction differs from $L(M)$ by a logical operator, causing the teleported output block to inherit a logical error.

\subsection{Logical State Preparation}
\label{sec:circ-state-prep}

The error correction by teleportation of Sec.~\ref{sec:circ-ect} consumes an encoded Bell pair. This pair is built by a noisy circuit from two logical-zero blocks. The circuit prepares two level-$L$ blocks in the logical-zero state, applies a transversal logical Hadamard to one block to obtain the logical-plus state, and applies a transversal logical CNOT between the two. The physical error model of Sec.~\ref{sec:circ-noise} acts throughout these preparations, gates, and measurements. A plain unitary encoding circuit is therefore not fault tolerant. A single faulty CNOT is fanned out by the remaining encoding CNOTs into a correlated error, which can exceed the correction capability of the code. If the Bell resource carries such an error into ECT, it either corrupts the teleportation measurements or passes directly to the teleported block. This produces a logical error at rate $O(p)$. Fault-tolerant state preparation therefore verifies each candidate block and rejects these faults before the block enters a Bell resource.

\begin{figure}[!htbp]
    \centering
    \includegraphics[width=\linewidth,height=0.8\textheight,keepaspectratio]{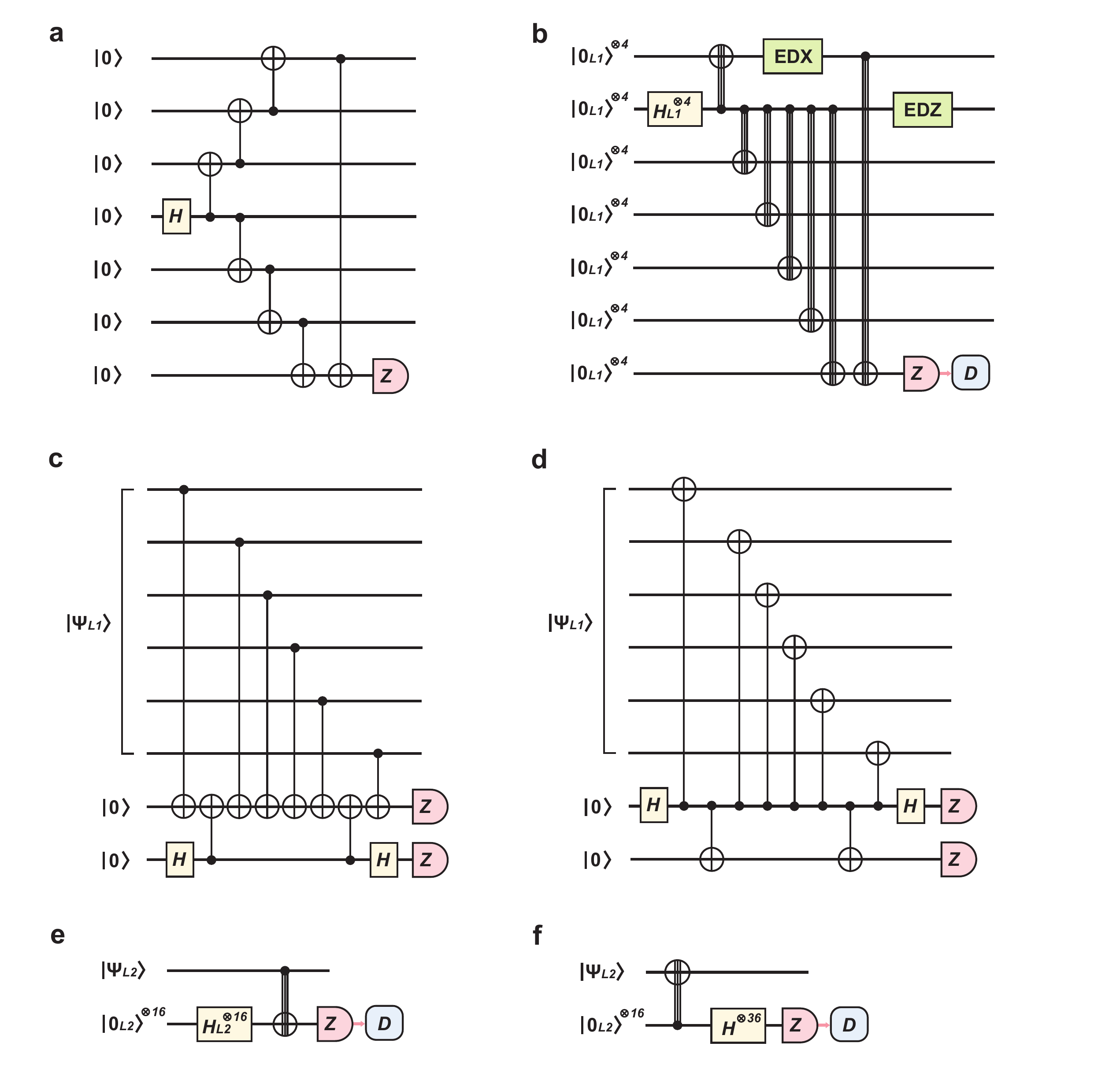}
    \caption{Fault-tolerant logical-zero state preparation for the many-hypercube codes, adapted from Fig.~5 of Ref.~\cite{goto2024high}. (a) Flagged level-1 encoder, which prepares the six-qubit GHZ state with the seventh qubit as a verification ancilla measured in the $Z$ basis. (b) Level-2 encoder, which combines seven accepted level-1 blocks into a level-2 GHZ state, applies the $X$- and $Z$-error-detection gadgets (EDX and EDZ) for error detection, and measures a seventh block as a verification ancilla. The level-3 encoder is not drawn separately and follows the same architecture, with the level-2 EDX and EDZ gadgets replacing the level-1 ones. (c) Flagged level-1 EDX gadget, which copies $X$ errors onto a $Z$-basis ancilla while a flag qubit catches faults on the ancilla. (d) Flagged level-1 EDZ gadget, which collects $Z$ errors on an ancilla through CNOTs while a flag qubit catches faults on the ancilla. (e) Level-2 EDX gadget, in which a transversal CNOT couples the level-2 block to a fresh encoded ancilla prepared in the logical-plus state, measured in the $Z$ basis through a decoder. (f) Level-2 EDZ gadget, in which a fresh encoded ancilla prepared in the logical-zero state controls a transversal CNOT into the level-2 block, collecting $Z$ errors, and is measured in the logical $X$ basis through a decoder. A rejected candidate is discarded and the preparation restarts.}
    \label{fig:goto-state-prep}
\end{figure}

The zero-state preparation of Ref.~\cite{goto2024high} uses a recursive prepare--detect--postselect procedure. Each level is built from accepted blocks of the previous level. A candidate is rejected whenever a verification step detects an error. The level-1 to level-3 encoders are shown in Fig.~\ref{fig:goto-state-prep}. A level-1 block is accepted only if its verification ancilla is measured to zero in the $Z$ basis.

A level-2 block is accepted only if neither error-detection gadget detects an error and the decoded $Z$-basis outcome of its verification block vanishes. The level-3 encoder is not drawn separately. It raises the level-2 construction by one level, using the level-2 error-detection gadgets in place of the level-1 gadgets.

\begin{figure}[!htbp]
    \centering
    \includegraphics[width=\linewidth]{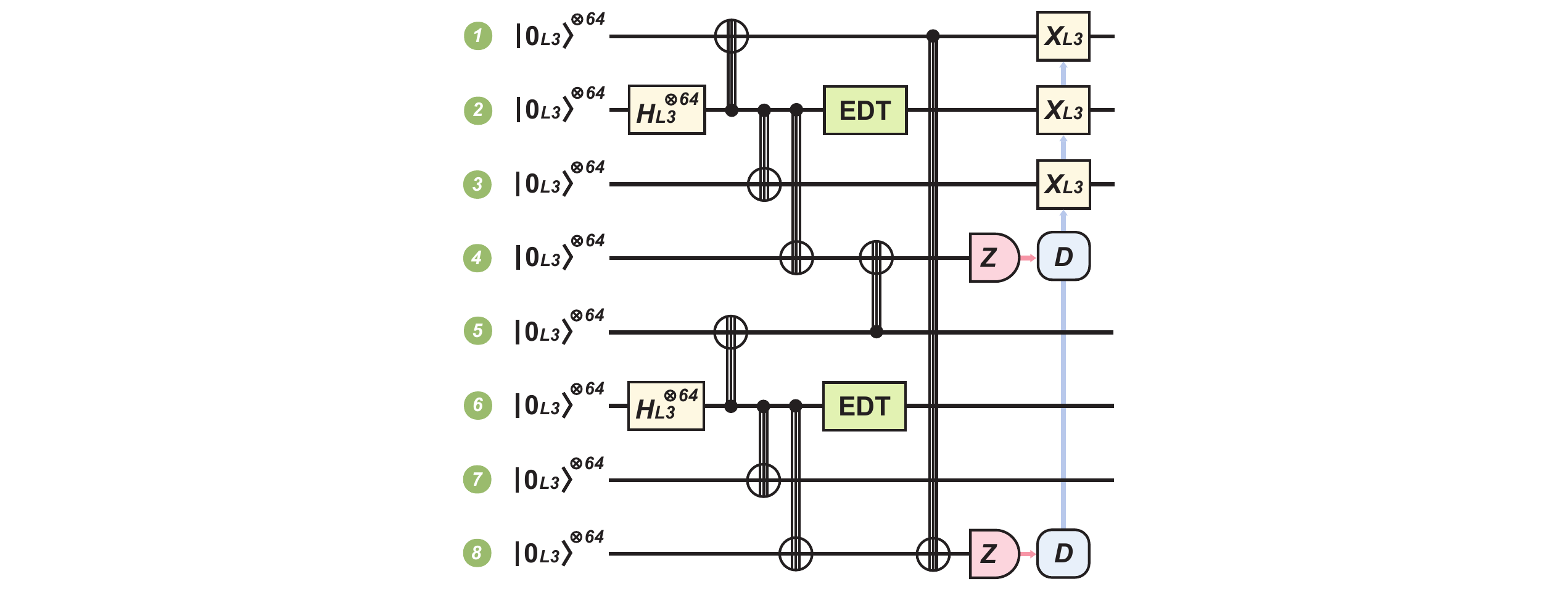}
    \caption{Level-4 logical-zero-state encoder, adapted from Fig.~5G of Ref.~\cite{goto2024high}. Eight accepted level-3 blocks, labeled $1$--$8$, enter in the logical-zero state. Logical Hadamards act on blocks $2$ and $6$, and transversal CNOTs fan out from block $2$ to blocks $1$, $3$, $4$ and from block $6$ to blocks $5$, $7$, $8$, forming two four-block level-3 GHZ resources; error detection by teleportation (EDT) is applied to the control blocks $2$ and $6$. The transversal CNOTs $5\to4$ and $1\to8$ then combine the two resources into a six-block GHZ state, and blocks $4$ and $8$ are measured in the $Z$ basis through decoders. If the two decoded outcomes agree, the encoded $X_{L3}$ correction determined by the common outcome is applied to blocks $1$, $2$, and $3$, and the surviving blocks $\{1,2,3,5,6,7\}$ form the level-4 block; otherwise the candidate is discarded and the preparation restarts.}
    \label{fig:level4-encoder}
\end{figure}

Level-4 preparation with full postselection on any detected error gives a very low acceptance probability. Consequently, level 4 uses a relaxed postselection rule. The level-4 encoder is shown in Fig.~\ref{fig:level4-encoder}.

The relaxed postselection rule of Ref.~\cite{goto2024high} uses error detection by teleportation (EDT) to evaluate a level-3 block. In this step, the Knill teleportation of Sec.~\ref{sec:circ-ect} measures the syndrome and decodes a candidate recovery. The rule then accepts the block under two conditions (labeled (C2) and (C3) in Ref.~\cite{goto2024high}):
\begin{enumerate}
    \item[(C2)] Each of the six level-2 sub-blocks has a unique minimum-weight candidate recovery.
    \item[(C3)] The whole level-3 block has a unique minimum-weight candidate recovery.
\end{enumerate}

The scheme of Ref.~\cite{goto2024high} verifies both conditions with a full level-by-level minimum-distance decoder that sums distances across sub-blocks. Our simulation uses a slightly stricter rule. It accepts a level-3 block in EDT under two different conditions:
\begin{enumerate}
    \item[(C2')] Each of the six level-2 sub-blocks has a minimum-weight candidate recovery with weight at most 1.
    \item[(C3')] The six level-2 sub-block corrections, stacked together, are consistent with the level-3 syndrome.
\end{enumerate}

Condition (C2$'$) is equivalent to condition (C2). The level-2 many-hypercube code has distance 4, so only errors of weight at most 1 can have a unique minimum-weight recovery. Conditions (C3$'$) and (C3) are related but not equivalent. Suppose the six minimum-weight sub-block recoveries, stacked together, are consistent with the level-3 syndrome. The whole level-3 block then necessarily has a unique minimum-weight candidate recovery. Hence, condition (C3$'$) implies condition (C3). The converse fails. The full level-by-level minimum-distance decoder can have a unique minimum-weight candidate even when the stacked per-sub-block recoveries are not globally consistent. Condition (C3$'$) is therefore strictly stronger than condition (C3). In exchange, our stricter rule admits a fast level-2 implementation. Each sub-block is decoded through a precomputed lookup table indexed by its level-2 syndrome, which avoids the full level-by-level minimum-distance decoder.

This relaxed EDT acceptance rule is not distance-preserving. A weight-6 incoming error can be accepted even though the accepted block is no longer in the logical-zero state. To see how this happens, consider one of the two level-3 blocks checked by EDT, such as the control blocks $2$ and $6$. Let $L_X^{(3)}$ be a minimum-weight logical-$X$ operator on that level-3 block. This operator has a weight of $2^3=8$. We can choose a representative for $L_X^{(3)}$ that is supported with weight $4$ on each of two level-2 sub-blocks. Next, we construct a smaller operator $R$ by taking exactly one physical factor of $L_X^{(3)}$ from each of these two sub-blocks. This gives $R$ a weight of $2$.

Now suppose the incoming error is
\begin{equation}
    E=L_X^{(3)}R,\qquad \operatorname{wt}(E)=6.
\end{equation}
Because the two factors in $R$ are shared with $L_X^{(3)}$, they cancel out in the product. The resulting error $E$ has weight $3$ in each of the two affected sub-blocks and acts trivially everywhere else. Furthermore, the logical operator $L_X^{(3)}$ has a vanishing syndrome, which leaves $\sigma(E)=\sigma(R)$. The exact level-2 lookup decoders see this syndrome and return the weight-1 recovery $R$ in each affected sub-block. This satisfies both condition (C2$'$) and condition (C3$'$), so the EDT step accepts the block. The EDT procedure records $R$ as a Pauli-frame update rather than applying it physically. Relative to this updated frame, the accepted block is left with the residual error $ER=L_X^{(3)}$, which is a weight-8 logical-$X$ error. Thus, six physical errors suffice to trick the system into accepting a level-3 block that already carries a full logical error. This corrupted block is then used as a logical-zero input by the level-4 encoder of Fig.~\ref{fig:level4-encoder}.

Our implementation realizes this postselection by maintaining a pool of accepted logical-zero blocks at each level. A candidate enters the pool only when it passes the detection step for its respective level. The pool is continuously replenished until it holds enough accepted blocks to build the noisy Bell resources consumed by ECT. A rejected candidate never enters a Bell resource. Its rejection incurs overhead in physical preparation, gates, and measurements, but it contributes nothing to the logical error event. An accepted block is not entirely error-free, however. It carries the residual error from its preparation. Before the block enters the pool, this residual error is reduced to a canonical form, as described in Sec.~\ref{sec:circ-dataset}.

\subsection{Logical CNOTs as the Benchmark}
\label{sec:circ-cnot-benchmark}

To evaluate a decoder on the full circuit-level benchmark, we follow the protected transversal logical CNOT simulation of Ref.~\cite{goto2024high}. Because a level-$L$ many-hypercube block encodes
\begin{equation}
    K^{(L)}=k^L=4^L
\end{equation}
logical qubits, applying physical CNOTs transversally between two such blocks implements $K^{(L)}$ logical CNOTs simultaneously. To prevent errors from accumulating unchecked, each transversal gate is immediately followed by ECT on both participating blocks, utilizing the fault-tolerantly prepared Bell resources described in Sec.~\ref{sec:circ-state-prep}. This combination of a noisy transversal gate and the subsequent ECT step constitutes a single protected CNOT round.

\begin{figure}[!htbp]
	\centering
	\includegraphics[width=1.0\linewidth]{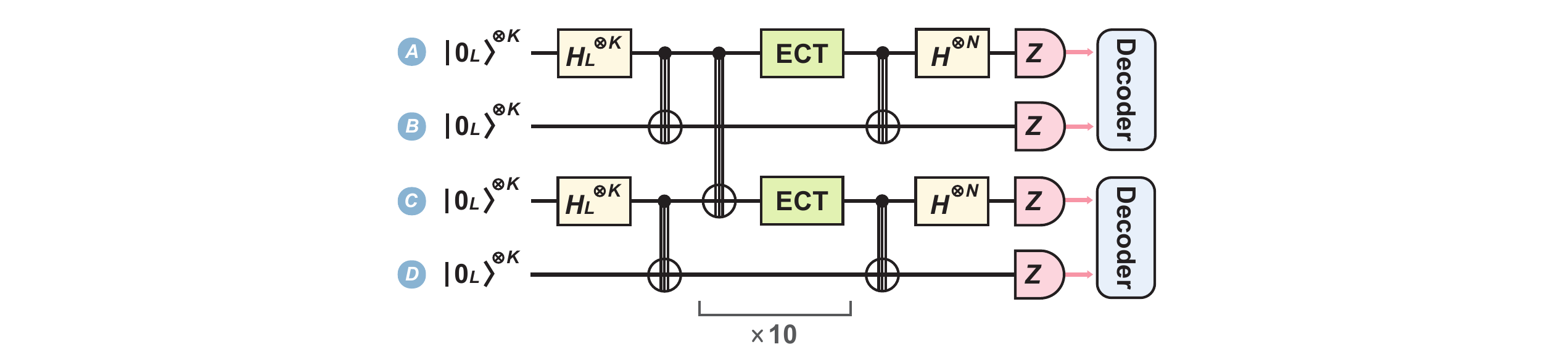}
	\caption{Ten-round logical-CNOT benchmark: blocks $A,B$ and $C,D$ are initialized as two ideal encoded Bell pairs; each round applies a noisy transversal CNOT $A\to C$ followed by ECT on $A$ and $C$. After ten rounds, an ideal Bell-state teardown and final measurement determine the logical outputs.}
	\label{fig:circuit}
\end{figure}

As illustrated in Fig.~\ref{fig:circuit}, the benchmark initializes four encoded blocks---labeled $(A,B)$ and $(C,D)$---as two ideal encoded Bell pairs. The protected operation is then repeated for $n_{\mathrm{round}}=10$ consecutive rounds: in each round, a noisy transversal CNOT couples block $A$ to block $C$, after which ECT is performed on blocks $A$ and $C$. Following these ten rounds, an ideal Bell-state teardown is executed, the blocks are ideally measured, and the final logical outputs are decoded. A trial is declared failed if any checked logical output is nonzero. While this pass/fail criterion matches Ref.~\cite{goto2024high}, our reporting metric differs: we report the logical CNOT failure rate per round for the whole logical block, rather than normalizing by the $K^{(L)}$ logical qubits contained within the block.

Let $n_{\mathrm{fail}}$ denote the number of failed trials among $n_{\mathrm{tot}}$ independent Monte Carlo samples. Following the notation of Ref.~\cite{goto2024high}, the estimated logical CNOT failure rate $P_{10}$ and its standard error $\Delta_{10}$ for the full ten-round experiment are given by
\begin{align}
    P_{10}&=\frac{n_{\mathrm{fail}}}{n_{\mathrm{tot}}}, \\
    \Delta_{10}&=\sqrt{\frac{P_{10}(1-P_{10})}{n_{\mathrm{tot}}}},
    \label{eq:circuit-p10-delta10}
\end{align}
where the uncertainty reflects the underlying binomial statistics of $n_{\mathrm{fail}}\sim\mathrm{Binomial}(n_{\mathrm{tot}},P_{10})$.

Assuming the failures are distributed evenly across the ten protected rounds, we convert the ten-round logical CNOT failure rate rate into a per-round logical CNOT failure rate $P_1$:
\begin{equation}
    P_1=1-(1-P_{10})^{1/n_{\mathrm{round}}},
    \qquad n_{\mathrm{round}}=10.
    \label{eq:circuit-p1-normalization}
\end{equation}
The per-round standard error $\Delta_1$ is then derived as:
\begin{equation}
    \Delta_1
    =\frac{\Delta_{10}}{n_{\mathrm{round}}}(1-P_{10})^{1/n_{\mathrm{round}}-1}.
    \label{eq:circuit-p1-delta1}
\end{equation}

\subsection{Generation of Circuit-Level Datasets}
\label{sec:circ-dataset}

The training data are generated inside the circuit-level benchmark itself, as a byproduct of the simulation. At every ECT step in ten-round logical CNOT benchmark, the simulation obtains the noisy outcome $M'=M\oplus E$ of Sec.~\ref{sec:circ-ect}. Each of the $2n$ bits of $M'$ is then flipped independently with probability $p$ to reproduce the noisy measurement. From this symplectic vector the simulation reads the decoder input, the syndrome $\sigma(E)=\sigma(M')$, and from the exact tracked error $E$ it reads the training target, the logical error class $L(E)$. Because the simulation tracks the circuit error $E$ directly, the target $L(E)$ is computed from the exact stabilizer and logical parity maps of the code, never from a decoder. The resulting dataset is independent of any decoder, separating the training data from the decoding performance.

Each ECT step thus supplies one training sample. The data are partitioned by physical error rate $p$ into shards of $1024$ circuit trajectories; each trajectory contributes $20$ intermediate ECT error events (one per ECT step over the ten rounds and two protected blocks) plus $2$ final-readout error events, yielding $22$ samples in total. For fine-tuning we use training rates $p_{\mathrm{train}}\in\{0.002,0.003,0.004,0.005\}$ at $8$ shards ($8192$ trajectories) per rate, and one validation rate $p_{\mathrm{valid}}=0.005$ at $3$ shards ($3072$ trajectories).

Although the target $L(E)$ is known exactly, it must be extracted with care due to ambiguities in Pauli-frame representation for prepared states. For a prepared logical-zero block, the operators that act trivially on the state include not only the code stabilizers but also all logical-$Z$ operators. Because the simulation strictly tracks Pauli frames, a physically trivial error (such as a residual $Z$ component from a faulty CNOT gate during preparation) can accumulate as a spurious logical-$Z$ operator in the tracked error $E$.

When this block is subsequently used to construct the encoded Bell pair $|\mathrm{Bell}_L\rangle$ (composed of halves 1 and 2) for ECT (Sec.~\ref{sec:circ-ect}), this spurious logical operator is carried over. It evolves into a logical operator that stabilizes the Bell state, manifesting as a spurious $X_L^{(1)} \otimes X_L^{(2)}$ or $Z_L^{(1)} \otimes Z_L^{(2)}$ error in the Pauli frame (depending on whether the original zero block served as the Hadamard-transformed control or the target in the Bell construction). Physically, these operators are perfectly harmless because they are seamlessly canceled by the ECT measurement mechanism. For instance, consider a spurious $X_L^{(1)} \otimes X_L^{(2)}$ error. The $X_L^{(1)}$ component flips the logical measurement outcome of half 1. This altered measurement instructs the ECT scheme to apply an additional $X_L$ correction to the surviving half 2, which exactly cancels the remaining $X_L^{(2)}$ component. A similar cancellation occurs for $Z_L^{(1)} \otimes Z_L^{(2)}$: the $Z_L^{(1)}$ component propagates to the input data block, flips its measurement outcome, and triggers a $Z_L$ correction that perfectly cancels $Z_L^{(2)}$.

However, while the physical state is flawlessly protected, the simulator's bookkeeping is altered. Because the logical measurement outcomes are flipped, the decoded logical label $L(E)$ absorbs an artificial logical $X$ or $Z$ shift. Meanwhile, the syndrome $\sigma(E)$ remains completely unchanged, as logical operators yield a vanishing syndrome. This arbitrary bookkeeping artifact mapping physically identical states to different logical labels exacerbates the inherent degeneracy of the code, polluting the conditional distribution $P(L(E)\mid\sigma(E))$ used for training.

To resolve this ambiguity, we apply a residual-error reduction. Once a logical-zero block has been successfully prepared (Sec.~\ref{sec:circ-state-prep}), its tracked error $E$ is replaced by a minimum-weight representative $E'$ that produces the same syndrome, $\sigma(E) = \sigma(E')$. Crucially, this reduction is performed differently for the $Z$ and $X$ sectors due to the asymmetry of the $|0_L\rangle$ state. For $Z$ errors, both the $Z$-type stabilizers and the logical-$Z$ operators act trivially on $|0_L\rangle$ (yielding a $+1$ eigenstate) and produce zero syndrome. Thus, at level 2 we invoke the exact lookup-table minimum-weight decoder strictly on the $Z$ sector: we compute the syndrome of the tracked $Z$-error $E_Z$, and use the decoder to directly find the minimum-weight error $E'_Z$ matching this syndrome, which then entirely replaces $E_Z$. For level 1, simple manual reduction rules are applied to both $X$ and $Z$ sectors based on stabilizer weights. Conversely, for $X$ errors at level 2, only $X$-type stabilizers act trivially (since a logical-$X$ would flip the state to $|1_L\rangle$), so we apply no such reduction to the $X$ sector beyond standard stabilizer equivalence. At higher levels (levels 3 and 4), we do not explicitly perform this same-syndrome frame replacement, and instead rely purely on the error-detection postselection of conditions (C2$'$) and (C3$'$) (Sec.~\ref{sec:circ-state-prep}) to reject heavily corrupted frames. By stripping this stabilizer dressing where applicable, we reduce the label pollution caused by simulation bookkeeping, allowing the observed $P(L(E)\mid\sigma(E))$ to more closely reflect the genuine degeneracy of the code.

\subsection{Fine-Tuning the Neural Decoder}

The code-capacity decoder of Sec.~\ref{app:decoder} is pretrained on code-capacity data, where errors are independent and identically distributed on the physical qubits. The circuit-level dataset instead contains correlated errors, and the error probability of a qubit depends on its position in the concatenation hierarchy. To capture this, we replace the fixed random initial state of the pretrained decoder,
$\mathbf h_q^{(0)}\sim\operatorname{Unif}([0,0.01]^r)$ in Sec.~\ref{app:decoder}, with a learnable initial embedding: each variable node $q=P_\lambda^{(\ell)}$ at every level $\ell$ now carries a learned vector $\boldsymbol\theta_q\in\mathbb R^r$ as its initial state, $\mathbf h_q^{(0)}=\boldsymbol\theta_q$, initialized from $\mathcal N(0,0.01)$ and trained jointly with the rest of the network. This lets the decoder store the level- and position-dependent error distribution of the circuit.

Apart from this change, the fine-tuning process is the same as the code-capacity training of Sec.~\ref{app:decoder}: the same $\mathcal L=\mathcal L_{\mathrm{phy}} + \mathcal L_{\mathrm{log}}$ objective, the same Muon--AdamW optimizer, and the same data exposure per epoch. The only differences are (i) the recurrent stack is unrolled $R=7$ times during training instead of $R=5$; (ii) the learning rate is one tenth of the code-capacity value (Muon $10^{-4}$, AdamW $10^{-5}$); and (iii) the training data are the circuit-level samples of Sec.~\ref{sec:circ-dataset} instead of fresh
code-capacity samples, drawn with a smaller batch size $20$ but the same total number of training samples per epoch. The training samples in each epoch contain $90\%$ intermediate ECT error events and $10\%$ final-readout error events. Fine-tuning runs for only $20$ epochs, far fewer than the $500$ epochs of the code-capacity training; this short schedule, together with the smaller circuit-level dataset that is reused within every epoch, avoids overfitting and preserves the generalization of the pretrained decoder. The selected decoder is evaluated in the full end-to-end circuit-level benchmark of Sec.~\ref{sec:circ-cnot-benchmark}, where it acts inside the error-correction loop and its residual errors propagate through later rounds.

\end{document}